\documentclass{emulateapj}

\shorttitle{A new RRL maser object toward MonR2-IRS2}
\shortauthors{Jim\'enez-Serra et al.}

\begin{document}

\title{A new radio recombination line maser object toward the MonR2 HII region}

\author{I. Jim\'{e}nez-Serra\altaffilmark{1}, A. B\'aez-Rubio\altaffilmark{2}, V. M. Rivilla\altaffilmark{2}, J. Mart\'{\i}n-Pintado\altaffilmark{2}, Q. Zhang\altaffilmark{1}, M. Dierickx\altaffilmark{1} and N. Patel\altaffilmark{1}}

\altaffiltext{1}{Harvard-Smithsonian Center for Astrophysics, 
60 Garden St., Cambridge, MA 02138, USA; ijimenez-serra@cfa.harvard.edu, qzhang@cfa.harvard.edu, mdierickx@cfa.harvard.edu, npatel@cfa.harvard.edu}

\altaffiltext{2}{Centro de Astrobiolog\'{\i}a (CSIC/INTA),
Ctra. de Torrej\'on a Ajalvir km 4,
E-28850 Torrej\'on de Ardoz, Madrid, Spain; 
ryvendel@gmail.com, jmartin@cab.inta-csic.es, baezra@cab.inta-csic.es}

\begin{abstract}

We report the detection of a new radio recombination line (RRL) maser object toward the IRS2 source in the MonR2 ultracompact HII region. The continuum emission at 1.3$\,$mm and 0.85$\,$mm and the H30$\alpha$ and H26$\alpha$ lines were observed with the Submillimeter Array (SMA) at angular resolutions of $\sim$0.5$"$-3$"$. The SMA observations show that the MonR2-IRS2 source is very compact and remains unresolved at spatial scales $\leq$400$\,$AU. Its continuum power spectrum at millimeter wavelengths is almost flat ($\alpha$=-0.16, with $S_\nu$$\propto$$\nu^\alpha$), indicating that this source is dominated by optically thin free-free emission. The H30$\alpha$ and H26$\alpha$ RRL emission is also compact and peaks toward the position of the MonR2-IRS2 source. The measured RRL profiles are double-peaked with the H26$\alpha$ line showing a clear asymmetry in its spectrum. Since the derived line-to-continuum flux ratios ($\sim$80 and 180$\,$km$\,$s$^{-1}$ for H30$\alpha$ and H26$\alpha$, respectively) exceed the LTE predictions, the RRLs toward MonR2-IRS2 are affected by maser amplification. The amplification factors are however smaller than those found toward the emission line star MWC349A, indicating that MonR2-IRS2 is a weakly amplified maser. Radiative transfer modelling of the RRL emission toward this source shows that the RRL masers arise from a dense and collimated jet embedded in a cylindrical ionized wind, oriented nearly along the direction of the line-of-sight. High-angular resolution observations at sub-millimeter wavelengths are needed to unveil weakly amplified RRL masers in very young massive stars.  

\end{abstract}


\section{Introduction}

Calculations of the level populations of atomic hydrogen in HII regions have shown that global population inversions can exist across the Rydberg levels of the hydrogen atom\footnote{The spontaneous decay rate decreases with quantum number $n$ as $n^{-5}$, leading to an underpopulation of the lower-$n$ levels of the hydrogen atom \citep[see e.g.][]{str96}.} \citep{cil36,bak38}. Population inversions lead to the formation of recombination lines under non-LTE conditions so that stimulated emission can occur. The general understanding is that most HII regions in our Galaxy are optically thin at wavelengths $\geq$1$\,$cm and stimulated emission is practically negligible \citep{wal90}. However, toward some ultracompact (UC) HII regions, the presence of ionized stellar winds modifies the internal electron density structure of these sources (with $n_e$$\geq$10$^7$$\,$cm$^{-3}$), leading to the formation of optically thick cores where radio recombination line (RRL) masers can form \citep{mar89,mar94}. RRL maser amplification is therefore expected to be a common phenomenon in UC HII regions with evidence of stellar winds \citep{mar02}.

The first RRL maser object was discovered by \citet{mar89} toward the emission line star MWC349A. This source has an edge-on disk and a bipolar ionized flow \citep[][]{coh85,pla92,wei08,mar11}, whose continuum power law spectrum \citep[$\alpha$$\sim$0.6, with $S_\nu$$\propto$$\nu^{\alpha}$;][]{alt81} is characteristic of a constant velocity stellar wind \citep{oln75}. \citet{mar89} showed that while the RRLs at 3$\,$mm are faint with single Gaussian profiles, the RRLs at 1.3$\,$mm are double-peaked with intensities factors of $\geq$50 brighter than the lines at 3$\,$mm. Since all RRLs with $\lambda$$\leq$2$\,$mm are largely amplified \citep{thu98}, \citet{mar89} concluded that the RRLs toward MWC349A are masers. It has been proposed that $\eta$ Carinae and Cepheus A HW2 also show RRL maser emission \citep[][]{cox95,jim11}. However, MWC349A still remains as the only RRL maser object firmly detected to date.

In this Letter, we report the detection of a new RRL maser object toward the Monoceros R2 (MonR2) UC HII region with the SMA \citep[$d$$\sim$830$\,$pc;][]{her76}. MonR2 is a {\it blister-type} HII region \citep[diameter of $\sim$27$"$;][]{mas85,woo89} that hosts a cluster of IR sources \citep[e.g.][]{car97}. Among these sources, IRS2 is a compact Young Stellar Object \citep[YSO;][]{alv04} with a luminosity of $\sim$5000$\,$L$_\odot$ \citep{how94}. This source (hereafter, MonR2-IRS2) is responsible for the spherical reflection nebula reported by \citet{asp90} in the MonR2 UC HII region. The detection of strong blue-shifted asymmetries in the H26$\alpha$ RRL toward this source by the SMA, reveals that RRL maser amplification can form in dense ionized winds toward very young massive stars.   

\section{Observations}
\label{obs}

%TABLE1------------------------------------------------------------------
\begin{deluxetable*}{lccccccccc}
\tabletypesize{\scriptsize}
%\rotate
\tablecaption{Instrumental parameters of the SMA observations.\label{tab1}}
\tablewidth{0pt}
\tablehead{
\colhead{Date} & \colhead{Config.} & \colhead{Line} & \colhead{LO Freq.} & \colhead{Synthesized Beam} & \colhead{$\tau_{225GHz}$} & \colhead{T$_{sys}$} & \colhead{BP Cal.} & \colhead{Flux Cal.} & \colhead{Gain Cal.} \\
\colhead{} & \colhead{} & \colhead{} & \colhead{(GHz)} & \colhead{($"$$\times$$"$, P.A.)} & \colhead{} & \colhead{(K)} & \colhead{} & \colhead{} & \colhead{}}
% & & & \colhead{(GHz)} & & & & (K) & }
\startdata

2010 Feb 13 & VEX & H30$\alpha$ & 224.611 & 0.53$"$$\times$0.37$"$, 48$^\circ$ & 0.05 & 200-240 & 3C273 & Titan & 0607-085/0730-116 \\ 

2010 Feb 20 & VEX & H30$\alpha$ & 224.611 & 0.53$"$$\times$0.37$"$, 48$^\circ$ & 0.03 & 200-240 & 3C273 & Titan & 0607-085/0730-116 \\ 

2011 Nov 15 & COM & H30$\alpha$ & 226.227 & 2.75$"$$\times$2.73$"$, -77$^\circ$ & 0.07 & 200-240 & BLLAC & Ganymede & 0607-085/0530+135 \\

2011 Nov 15 & COM & H26$\alpha$ & 348.324 & 2.29$"$$\times$1.50$"$, -39$^\circ$ & 0.07 & 400-500 & BLLAC & Ganymede & 0607-085/0530+135

\enddata

\end{deluxetable*}
%------------------------------------------------------------------

Observations of the H30$\alpha$ line (231.9$\,$GHz) toward the MonR2-IRS2 source were carried out with the SMA\footnote{The Submillimeter Array is a joint project between the Smithsonian Astrophysical Observatory and the Academia Sinica Institute of Astronomy and Astrophysics, and is funded by the Smithsonian Institution and the Academia Sinica.} in the very extended (VEX) configuration in two tracks in single-receiver mode (4$\,$GHz bandwidth per sideband). In addition, the H30$\alpha$ and the H26$\alpha$ (353.6$\,$GHz) RRLs were simultaneously observed in a third track in compact (COM) configuration in dual-receiver mode, which provided 2$\,$GHz of total bandwidth per sideband and receiver. The instrumental parameters of the SMA observations are reported in Table$\,$\ref{tab1}. The phase center of the observations was set at $\alpha(J2000)$=06$^{h}$07$^{m}$45.83$^s$, $\delta(J2000)$=-06$^{\circ}$22$'$53.50$''$. We used a uniform spectral resolution of 0.8$\,$MHz, which provided a velocity resolution of $\sim$1.1$\,$km$\,$s$^{-1}$ at 231.9$\,$GHz, and of $\sim$0.7$\,$km$\,$s$^{-1}$ at 353.6$\,$GHz. Data calibration was carried out within the IDL MIR software package, and continuum subtraction, imaging and deconvolution was done within MIRIAD. The uncertainty in the flux calibration was within 20\%. 
 
\section{Results}
\label{res}

%TABLE2------------------------------------------------------------------
\begin{deluxetable*}{lccccc}
\tabletypesize{\scriptsize}
%\rotate
\tablecaption{Derived parameters of the continuum and RRL emission toward MonR2-IRS2.\label{tab2}}
\tablewidth{0pt}
\tablehead{
& \multicolumn{5}{c}{Continuum Emission} \\ 
Config. & $\lambda$ & \multicolumn{2}{c}{Position Cont. Peak} & \colhead{Peak Flux\tablenotemark{a}} & \colhead{Angular Size} \\
& (mm) & \colhead{$\alpha$(J2000)} & \colhead{$\delta$(J2000)} & \colhead{(Jy beam$^{-1}$)} & \colhead{($"$$\times$$"$, P.A.)}}
\startdata

COM & 1.3 & 06$^h$07$^m$45.804$^s$ & -06$^\circ$22$'$53.50$''$ & 0.154$\pm$0.005 & 3.13$"$$\times$2.79$"$, 105$^\circ$ \\ 

VEX & 1.3 & 06$^h$07$^m$45.806$^s$ & -06$^\circ$22$'$53.53$''$ & 0.157$\pm$0.002 & 0.55$"$$\times$0.38$"$, 45$^\circ$ \\ 

COM & 0.85 & 06$^h$07$^m$45.807$^s$ & -06$^\circ$22$'$53.45$''$ & 0.14$\pm$0.01 & 2.44$"$$\times$1.60$"$, 130$^\circ$  \\ \hline \hline

& \multicolumn{5}{c}{RRL Emission} \\
Config. & \colhead{Line} & \colhead{$T_L \Delta v$\tablenotemark{b}} & \colhead{$v_{LSR}$} & \colhead{$\Delta v$} & \colhead{$T_L$\tablenotemark{c}} \\

& & \colhead{(Jy beam$^{-1}$ km$\,$s$^{-1}$)} & \colhead{(km$\,$s$^{-1}$)} & \colhead{(km$\,$s$^{-1}$)} & \colhead{(Jy beam$^{-1}$)} \\ \hline

COM & H30$\alpha$ & 6.93 (0.22) & -13.4 (0.6) & 23.4 (1.7) & 0.28 (0.03) \\
& &  8.36 (0.24) & 10.5 (0.0) & 30.0 (0.0) & 0.26 (0.03) \\
& &  4.69 (0.19) & 33.3 (0.6) & 18.9 (2.0) & 0.23 (0.03) \\

VEX & H30$\alpha$ & 4.2 (0.4) & -10.89 (0.12) & 15.4 (0.3) & 0.26 (0.07) \\
& & 4.7 (0.6) & 10.5 (0.0) & 30.0 (0.0) & 0.15 (0.07) \\
& & 3.3 (0.4) & 31.46 (0.15) & 13.7 (0.4) & 0.23 (0.07) \\

COM & H26$\alpha$ & 14.5 (0.5) & -14.0 (3.0) & 15.5 (3.0) & 0.88 (0.05) \\
& & 6.7 (0.7) & 10.5 (0.0) & 32.2 (3.0) & 0.20 (0.05) \\
& & 5.0 (0.4) & 28.9 (3.0) & 10.4 (3.0) & 0.45 (0.05) \\

\enddata

\tablenotetext{a}{The error in the peak flux corresponds to the 1$\sigma$ noise level in the continuum images.}
\tablenotetext{b}{The error in the integrated intensity flux of the RRLs is calculated as $\sigma_{Area}$=1$\sigma$$\times$$\sqrt{\delta v\times\Delta v}$, with $\delta v$ the velocity resolution in the RRL 
spectra (see Section$\,$\ref{res}), and $\Delta v$ the linewidth derived from the Gaussian fit of the RRL emission.}
\tablenotetext{c}{The error in the peak intensity of the H30$\alpha$ and H26$\alpha$ lines is given by the 1$\sigma$ noise level in the RRL images.}
\end{deluxetable*}
%------------------------------------------------------------------

%FIG1**************************************
\begin{figure}
\begin{center}
\includegraphics[angle=0,width=0.47\textwidth]{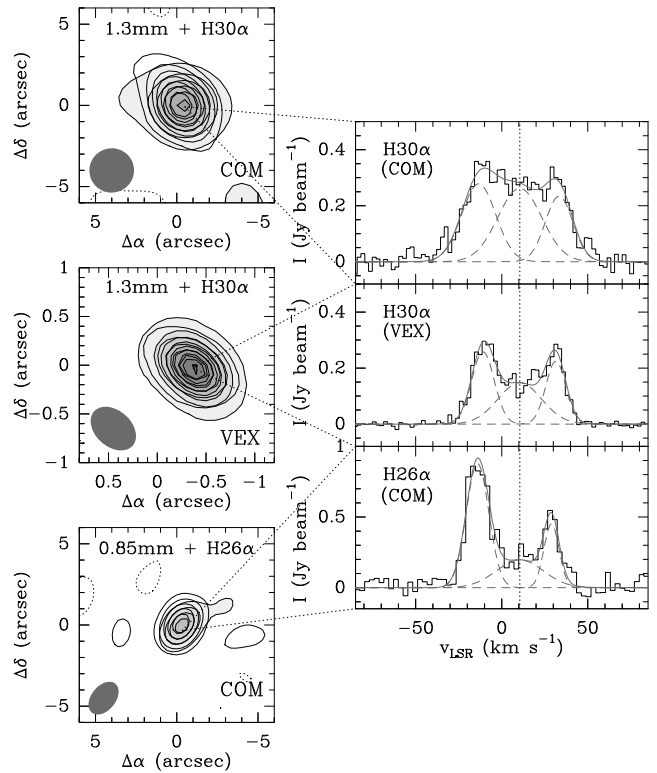}
\caption{{\it Left panels:} Integrated intensity images of the H30$\alpha$ line from -35.5 to 47.0$\,$km$\,$s$^{-1}$ (thick contours) observed with the SMA in COM and VEX (upper and middle panels), and of the H26$\alpha$ RRL measured toward MonR2-IRS2 in COM (lower panel). These images are superimposed on the continuum emission observed at 1.3$\,$mm in COM and VEX (grey scale; upper and middle panels) and at 0.85$\,$mm in COM (lower panel). Offsets are given in arcseconds with respect to the phase center of the observations. Note that the spatial scales shown for the VEX images (middle panel) are $\sim$5 times smaller than those of the COM data (upper and lower panels). The first contour and step level for the 1.3$\,$mm and H30$\alpha$ maps in COM (upper panel) are, respectively, 13.5 (3$\sigma$) and 27$\,$mJy$\,$beam$^{-1}$, and 2.4 (3$\sigma$) and 2.4$\,$Jy$\,$beam$^{-1}$$\,$km$\,$s$^{-1}$. For the 1.3$\,$mm and H30$\alpha$ maps in VEX (middle panel), the first contour and step level are 4.5 (3$\sigma$) and 18$\,$mJy$\,$beam$^{-1}$, and 0.75 (3$\sigma$) and 2.25$\,$Jy$\,$beam$^{-1}$$\,$km$\,$s$^{-1}$, respectively. For the 0.85$\,$mm and H26$\alpha$ images in COM (lower panel), these levels are 30 (3$\sigma$) and 30$\,$mJy$\,$beam$^{-1}$, and 4.5 (3$\sigma$) and 4.5$\,$Jy$\,$beam$^{-1}$$\,$km$\,$s$^{-1}$, respectively.
{\it Right panels:} Spectra of the H30$\alpha$ and H26$\alpha$ RRLs measured with the SMA in COM and VEX toward the MonR2-IRS2 continuum peak. Dashed curves show the individual Gaussian fits of the different velocity components detected toward MonR2-IRS2. Solid lines show the total Gaussian fit of the H30$\alpha$ and H26$\alpha$ RRLs. The vertical dotted line indicates the radial velocity of the source \citep[$v_{LSR}$=10.5$\,$km$\,$s$^{-1}$;][]{torr83}.}
\label{f1}
\end{center}
\end{figure}
%******************************************

In Figure$\,$\ref{f1} (left panels), we present the continuum images at 1.3$\,$mm measured toward MonR2-IRS2 in COM and VEX configurations (grey scale; upper and middle panels), and at 0.85$\,$mm in COM (lower panel). The derived parameters of this emission, obtained by performing 2D Gaussian fits, are given in Table$\,$\ref{tab2}. The position of the continuum peak is very close to that reported for the IRS2 source in the near-IR and X-rays \citep{car97,nak03}. The derived angular sizes for the MonR2-IRS2 source indicate that this object is very compact and unresolved in the SMA images (Table$\,$\ref{tab2}). Since the continuum peak intensity at 1.3$\,$mm is $\sim$0.15$\,$Jy$\,$beam$^{-1}$ in both VEX and COM, MonR2-IRS2 does not show any structure at sub-arcsecond scales (0.4$"$-0.5$"$ or 320-400$\,$AU). This is consistent with the results by \citet{alv04} from near-IR speckle imaging. By smoothing the 1.3$\,$mm and 0.85$\,$mm images in COM to the same angular resolution ($\sim$2.8$"$), we derive a decreasing spectral index of $\alpha$=$-$0.16 (with $S_\nu$$\propto$$\nu^\alpha$) for MonR2-IRS2. This spectral index is consistent with optically thin free-free continuum emission. We note that our estimate of the spectral index is not affected by missing flux because the MonR2-IRS2 source is very compact. Indeed, its measured flux at 1.3$\,$mm is the same in the VEX and COM beams (see Table$\,$\ref{tab2}), confirming the lack of large-scale structures in this source.

Superimposed on the continuum images, Figure$\,$\ref{f1} reports the integrated intensity emission of the H30$\alpha$ and H26$\alpha$ lines measured toward MonR2-IRS2 from $-$35.5$\,$km$\,$s$^{-1}$ to 47.0$\,$km$\,$s$^{-1}$ (thick contours). The RRL maps show a compact structure centered at the continuum peak, which is almost unresolved even in the VEX images at angular scales of $\sim$0.5$"$. The spectra of the H30$\alpha$ and H26$\alpha$ RRLs extracted from the position of the continuum peak are shown in the right panels of Figure$\,$\ref{f1}. These spectra have been smoothed to a velocity resolution of 2.2$\,$km$\,$s$^{-1}$ for H30$\alpha$, and of 2.8$\,$km$\,$s$^{-1}$ for H26$\alpha$. The H30$\alpha$ and H26$\alpha$ lines have bright double-peaked line profiles whose components are red- and blue-shifted by $\sim$20-25$\,$km$\,$s$^{-1}$ with respect to the radial velocity of the source \citep[$v_{LSR}$$\sim$10.5$\,$km$\,$s$^{-1}$;][]{torr83}. In addition, some faint emission at $v_{LSR}$$\sim$10.5$\,$km$\,$s$^{-1}$ is also detected, likely associated with the extended MonR2 UC HII region. This would explain why the double-peaked line profile of H30$\alpha$ is not as clearly seen in COM as in VEX, since the contribution from the extended emission would be larger within the COM beam. 

From Figure$\,$\ref{f1}, we also find that while the double-peaked H30$\alpha$ RRLs have similar peak intensities of $\sim$0.2-0.3$\,$Jy$\,$beam$^{-1}$, the H26$\alpha$ line profile shows a clear asymmetry with the blue-shifted gas being factors of $\sim$2 brighter than the red-shifted emission ($\sim$0.9$\,$Jy$\,$beam$^{-1}$ vs. $\sim$0.5$\,$Jy$\,$beam$^{-1}$, respectively). This behavior can only be explained if stimulated emission plays a key role in the formation of the H26$\alpha$ RRL \citep[see][]{mar93,jim11,bae12}. As shown in Section$\,$\ref{mod}, the H26$\alpha$ RRL blue-shifted asymmetry is a consequence of the presence of more background emission to be amplified by the blue-shifted foreground ionized gas than for the red-shifted material, assuming that the RRL masers are formed in an expanding ionized jet. 

The observed parameters of the three velocity components of the H30$\alpha$ and H26$\alpha$ RRLs are reported in Table$\,$\ref{tab2}. These parameters were derived by fixing the peak velocity and linewidth of the ionized component at ambient velocities to, respectively, 10.5$\,$km$\,$s$^{-1}$ (the radial velocity of the MonR2-IRS2 source) and to 30$\,$km$\,$s$^{-1}$ \citep[the expected thermal linewidth for RRLs in HII regions with electron temperatures of $T_e$=1-2$\times$10$^4$$\,$K; see e.g.][]{ket08}. The derived peak velocities of the H30$\alpha$ and H26$\alpha$ components are 29-33$\,$km$\,$s$^{-1}$ for the red-shifted gas, and $-$11 to $-$14$\,$km$\,$s$^{-1}$ for the blue-shifted emission. The measured linewidths are $\sim$10-15$\,$km$\,$s$^{-1}$ (Table$\,$\ref{tab2}), except for the H30$\alpha$ COM data whose linewidths are broader due to the contamination from the extended MonR2 UC HII region.

By integrating the emission of the H30$\alpha$ RRL from -18 to -4.8$\,$km$\,$s$^{-1}$ and from 27.1 to 38.1$\,$km$\,$s$^{-1}$, the VEX images of the blue-shifted and red-shifted H30$\alpha$ gas reveal a spatial shift of 0.045$"$ (or 360$\,$AU) in the southeast-northwest direction (blue-shifted emission toward the southeast, red-shifted gas toward the northwest). Although this shift is small compared to the angular resolution of the VEX data (0.5$"$$\times$0.4$"$; Table$\,$\ref{tab1}), higher positional accuracy of $\frac{\theta_{beam}}{2\times S/N}$ (with S/N the signal-to-noise ratio of the H30$\alpha$ emission) can be achieved thanks to the bright integrated intensities of the blue- and red-shifted H30$\alpha$ RRL components. By fitting 2D Gaussians to the integrated intensity images of these velocity components, we obtain an accuracy in the H30$\alpha$ Gaussian centroid position of $\sim$0.008$"$. This implies that the detected H30$\alpha$ spatial shift of 0.045$"$$\pm$0.008$"$ is at the 5.5$\sigma$ confidence level. This shift could be associated with a collimated ionized jet propagating at a velocity of 20$\,$km$\,$s$^{-1}$ (see Section$\,$\ref{mod}). 

\section{RRL Masers in MonR2-IRS2}
\label{ltr}

%TABLE3------------------------------------------------------------------
\begin{deluxetable}{lccc}
%\tabletypesize{\scriptsize}
%\rotate
\tablecaption{Integrated Line-to-Continuum flux ratios (ILTRs).\label{tab3}}
\tablewidth{0pt}
\tablehead{
\colhead{RRL} & \colhead{LTE\tablenotemark{a}} & \colhead{MonR2-IRS2\tablenotemark{b}} & \colhead{MWC349A} \\
& \colhead{(km$\,$s$^{-1}$)} & \colhead{(km$\,$s$^{-1}$)} & \colhead{(km$\,$s$^{-1}$)}}
\startdata

H30$\alpha$ & $\sim$65 & 80 & 298\tablenotemark{c} \\ 
H26$\alpha$ & $\sim$103 & 180 & 756\tablenotemark{d}

\enddata
%% Text for table notes should follow after the \enddata but before
%% the \end{deluxetable}. Make sure there is at least one \tablenotemark
%% in the table for each \tablenotetext.

\tablenotetext{a}{Calculated for optically thin continuum emission, $T_e^*$=10$^4$$\,$K and $N$(He$^+$)/$N$(H$^+$)=0.08.}
\tablenotetext{b}{Estimated from the ratio between the total area of the H30$\alpha$ and H26$\alpha$ lines and the continuum peak flux at 1.3$\,$mm and 0.85$\,$mm (Table$\,$\ref{tab2}).}
\tablenotetext{c}{From \citet{mar89}.}
\tablenotetext{d}{Derived from the average H26$\alpha$ line flux measured by \citet{thu94}, and from the expected free-free continuum flux at 0.85$\,$mm assuming a spectral index of $\alpha$$\sim$0.6 (Altenhoff et al. 1981).}

\end{deluxetable}
%------------------------------------------------------------------

The double-peaked RRL profiles with clear asymmetries in the low-$n$ RRL transitions toward MonR2-IRS2 (i.e. in the H26$\alpha$ line; Section$\,$\ref{res}), resemble the behaviour of the RRLs detected toward MWC349A, and suggest that the RRL emission toward MonR2-IRS2 at $\lambda$$\leq$1.3$\,$mm is affected by maser amplification. The integrated line-to-continuum flux ratios, ILTRs\footnote{ILTR is defined as $\Delta v T_L/T_C$, and depends on the RRL frequency as $\nu^{1.1}$ for LTE and optically thin emission \citep[see e.g.][]{mar89}.}, derived toward this source support this idea (Table$\,$\ref{tab3}). As in MWC349A \citep{mar89,thu94}, the ILTRs derived toward MonR2-IRS2 (80$\,$km$\,$s$^{-1}$ for H30$\alpha$ and 180$\,$km$\,$s$^{-1}$ for H26$\alpha$) exceed the values predicted under LTE conditions (65$\,$km$\,$s$^{-1}$ and 103$\,$km$\,$s$^{-1}$, respectively). However, unlike MWC349A, the RRL maser amplification factors toward MonR2-IRS2 with respect to LTE are significantly smaller than those found in MWC349A \citep[$\sim$1.2-1.7 for MonR2-IRS2 vs. $\geq$5-7 for MWC349A; Table$\,$\ref{tab3} and][]{str96,thu98}. This indicates that MonR2-IRS2 is a weakly amplified RRL maser. This is due to the fact that MWC349A has a larger optically thick core than MonR2-IRS2, allowing a much larger maser amplification of the RRLs \citep[see][and Section$\,$\ref{mod}]{pon94}.

\section{Modelling of the RRL Emission Toward MonR2-IRS2}
\label{mod}

%FIG2**************************************
\begin{figure}
\begin{center}
\includegraphics[angle=0,width=0.47\textwidth]{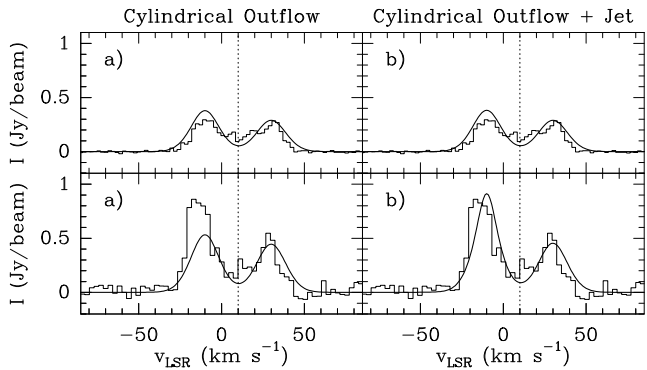}
\caption{Comparison of the H30$\alpha$ and H26$\alpha$ RRL spectra measured with the SMA toward MonR2-IRS2 (in VEX and COM, respectively), with those predicted by the MORELI code for two geometries of the MonR2-IRS2 source: a) a cylindrical, ionized outflow oriented along the line-of-sight (left panels); and b) a cylindrical, ionized outflow with an elongated and dense inhomogeneity resembling a collimated jet (right panels). Vertical dotted lines indicate the assumed radial velocity of the source $v_{LSR}$=10$\,$km$\,$s$^{-1}$. }   
\label{f2}
\end{center}
\end{figure}
%******************************************

By comparing the H30$\alpha$ and H26$\alpha$ RRLs measured toward MonR2-IRS2 with radiative transfer modelling of the RRL emission, we can constrain the physical structure and kinematics of the ionized gas toward this YSO. To do this, we have used the three-dimensional RRL radiative transfer code MORELI developed by \citet{mar11} and \citet{bae12}. In our calculations, we have used the departure coefficients, $b_n$ and $\beta_n$, derived by \citet{wal90}. For the geometry of the source, we have assumed that MonR2-IRS2 has a central mass of 13$\,$M$_\odot$, which is consistent with a B1 star on the ZAMS \citep[luminosity of $\sim$5000$\,$L$_\odot$;][]{pan73} and with the $Ly\alpha$ photon flux ($\sim$6$\times$10$^{45}$$\,$photons$\,$s$^{-1}$) derived from the 1.3$\,$mm and 0.85$\,$mm continuum data (Table$\,$\ref{tab2}). In addition, we consider that MonR2-IRS2 powers a cylindrical, isothermal ($T_e$=10$^4$$\,$K) ionized outflow, which expands at a velocity of 20$\,$km$\,$s$^{-1}$. The cylindrical outflow is defined so that its outer radius is 100$\,$AU and its electron density distribution $n_e$ decreases as $r^{-2}$, with $r$ the distance to the star and with $n_e$=10$^8$$\,$cm$^{-3}$ at a distance $r$$\sim$7$\,$AU. The total length of the cylinder is $L$=520$\,$AU and its axis is oriented along the direction of the line-of-sight. The assumed radial velocity for the MonR2-IRS2 source is $v_{LSR}$=10$\,$km$\,$s$^{-1}$. 

To reproduce the optically thin free-free continuum emission of MonR2-IRS2 at 1.3$\,$mm and 0.85$\,$mm, we consider a cavity around the star with radius $r$=13$\,$AU \citep[see][]{bae12}. The predicted continuum fluxes at 1.3$\,$mm and 0.85$\,$mm are 0.156$\,$Jy$\,$beam$^{-1}$ and 0.148$\,$Jy$\,$beam$^{-1}$, respectively, which differ from the observed values by less than 3\%.      

In Figure$\,$\ref{f2} (left panels), we compare the RRL profiles of the H30$\alpha$ and H26$\alpha$ lines measured with the SMA toward the MonR2-IRS2 source (in VEX and COM, respectively), with the synthetic line profiles predicted by the MORELI code for the cylindrical, ionized outflow assumed for MonR2-IRS2 (Case a). Although our model reproduces the double-peaked profile of the H30$\alpha$ RRL relatively well, it clearly fails to predict the asymmetry detected in the H26$\alpha$ line (Figure$\,$\ref{f2}). Indeed, the predicted line profiles for both the H30$\alpha$ and H26$\alpha$ RRLs are symmetric because the continuum emission of MonR2-IRS2 is optically thin throughout the ionized outflow, which prevents the strong amplification and the asymmetries of the RRLs.  

In order to explain the strongly asymmetric H26$\alpha$ line profile, larger electron densities are required for the inner regions of the ionized wind so that the continuum emission becomes locally optically thick. To simulate this, we have assumed two elongated inhomogeneities with $n_e$=10$^8$$\,$cm$^{-3}$ located at the axis of the blue-shifted and red-shifted lobes of the cylindrical ionized outflow, resembling a collimated ionized jet. The size of the inhomogeneities is 2$\,$AU$\times$2$\,$AU$\times$60$\,$AU, and they are placed on the axis of the cylindrical outflow at a distance of 43$\,$AU from the central star. 

In Figure$\,$\ref{f2} (right panels), we report the H30$\alpha$ and H26$\alpha$ line profiles predicted by our model with a cylindrical ionized outflow powered by a dense, collimated jet (Case b). While the predicted H30$\alpha$ emission shows a similar line profile to that obtained in Case a) for only a cylindrical ionized outflow, the model with the ionized outflow+jet perfectly matches the asymmetry observed in the H26$\alpha$ line profile. This is due to the fact that the inner regions of the jet are optically thick, allowing the maser amplification of the H26$\alpha$ line. Since the optically thick region in the plane of the sky of the ionized jet is very small (only 2$\,$AU$\times$2$\,$AU), its contribution to the total optical depth of the MonR2-IRS2 continuum source is practically negligible. We note that only the H26$\alpha$ RRL gets amplified because the departure coefficient $\beta_n$ of the H26$\alpha$ transition at an electron density of $n_e$=10$^8$$\,$cm$^{-3}$ is significantly larger than that of the H30$\alpha$ line \citep[see Figure$\,$8 in][]{str96}. The electron densities $n_e$ assumed for the elongated inhomogeneities do not likely exceed 10$^8$$\,$cm$^{-3}$, because the $\beta_n$ coefficients for both the H30$\alpha$ and H26$\alpha$ RRLs get close to zero, hindering the RRL maser amplification \citep[see Figure$\,$5 in][]{str96}.  

In our model for MonR2-IRS2 with an ionized outflow+jet (Case b), the elongation of the dense inhomogeneities is required to sufficiently amplify the blue-shifted peak of the H26$\alpha$ emission. In contrast with the blue-shifted peak, the RRL emission from the red-shifted lobe of the ionized jet does not get amplified because the optically thick continuum acts as a screen for the red-shifted H26$\alpha$ emission. 

Finally, we stress that other geometries for the MonR2-IRS2 source have been explored in our study. In the case of an edge-on Keplerian rotating disk with a bi-conical ionized flow \citep[similar to that assumed for MWC349A; see][]{mar11}, the model also predicts double-peaked line profiles for large opening angles of the ionized wind \citep[$\theta_a$$\geq$60$^\circ$; see][for the definition of this angle]{bae12}. However, the model not only fails to reproduce the blue-shifted asymmetry of the H26$\alpha$ line profile, but predicts a RRL emission excess at velocities close to the ambient cloud velocity that is not observed. The presence of high-density inhomogeneities could partially alleviate the discrepancies between the model predictions and the observations, but the best fit is obtained with the cylindrical ionized outflow powered by the elongated jet.

In summary, we report the detection of a new RRL maser object toward the IRS2 source in the MonR2 UC HII region. Our SMA images reveal that MonR2-IRS2 is a compact object with an spectral index $\alpha$=-0.16, characteristic of optically thin free-free emission. The line profiles of the H30$\alpha$ and H26$\alpha$ RRLs are double-peaked, and their derived ILTRs clearly exceed those predicted under LTE conditions. The RRL emission at $\lambda$$\leq$1.3$\,$mm toward the MonR2-IRS2 source are weakly amplified masers. Our radiative transfer modelling of the RRLs toward the MonR2-IRS2 source suggests that the RRL masers arise from a dense and highly collimated jet embedded in a ionized, cylindrical outflow, nearly oriented along the direction of the line-of-sight. Interferometric observations at sub-millimeter wavelengths have the potential to unveil a population of weakly amplified RRL maser objects in UC HII regions with high-velocity ionized winds.

\acknowledgments

We are grateful to the SMA director and staff for letting us carry out part of the SMA observations under the Harvard Astronomy Ay191 course, and to the Ay191 students for their enthusiasm during this course. We also acknowledge an anonymous referee for the constructive comments to the manuscript. I.J-S. acknowledges the Smithsonian Astrophysical Observatory for the support provided through a SMA fellowship. J.M.-P. and I.J.-S. have been partially funded by MICINN grants ESP2007-65812-C02-C01 and AYA2010-21697-C05-01 and AstroMadrid (CAM S2009/ESP-1496).

%{\it Facilities:} \facility{Submillimeter Array}.

\end{document}